# Visualization of out-of-plane spin generation in mirror symmetry broken Co

Yakun Liu,[1,b)] Fanrui Hu,[1,b)] Guoyi Shi,[1,2] and Hyunsoo Yang[1,2,a)]

[1]*Department of Electrical and Computer Engineering, National University of Singapore, Singapore 117576, Singapore*

[2]*NUS Graduate School for Integrative Sciences and Engineering, National University of Singapore, Singapore 119077, Singapore*

Generating out-of-plane spins in sputtered materials holds immense potential for achieving field-free spin-orbit torque switching in practical applications and mass production. In this work, we present the detection of out-of-plane spins from single-layer ferromagnetic Co layers, which are visualized through helicity-dependent photomapping techniques. Our experiments have shown that out-of-plane spin generation is dependent on the magnetization direction, current density, and Co thickness. Our findings indicate that amorphous sputtered Co can be a promising candidate as an out-of-plane spin source material for industrial massive production.

---

[a)] Author to whom correspondence should be addressed. Electronic mail: eleyang@nus.edu.sg.

[b)] Yakun Liu and Fanrui Hu contributed equally to this work.

Efficient and convenient field-free spin-orbit toque switching for perpendicular magnetic anisotropy (PMA) is critical for the widespread adoption of spintronic devices[1-14]. However, the effective generation of out-of-plane spin is a bottleneck for PMA switching. Conventional spin source materials such as heavy metals (Pt[15,16] or Ta[17], etc.) and topological insulators ($Bi_2Se_3$[18,19], etc.) predominantly facilitate in-plane spin generation. Materials with broken mirror symmetry have demonstrated potential for out-of-plane spin generation[2,5-8,12,13] with Weyl semimetals, such as $WTe_2$[2,3,5-7,12]. However, high-quality Weyl semimetals are primarily available in the single-crystal form, constraining their suitability for mass production. Recently, the $L1_1$-ordered PtCo alloy has emerged as a promising candidate for field-free PMA switching[8,20], but proper crystalline formation necessary for out-of-plane spin generation is highly dependent on annealing temperatures, which may degrade the adjacent commercial PMA layer. Alternative techniques, such as spin precession in a ferromagnet (FM)/normal metal (NM)[4], the planar Hall effect in asymmetry NM/FM/NM trilayer systems[21], lateral structure inversion asymmetry in contacts[22] and composition gradient in a ferrimagnet[23] have also been reported with out-of-plane spin generation. However, these methods require complex manufacturing processes which limit their practical applications. Thus, fabricating spin source structures that can be manufactured at room temperature via streamlined sputtered processes, is a desirable solution.

It has been predicted that the out-of-plane spin can also be generated in a single layer of ferromagnet, providing a new way to create out-of-plane spin and improve robustness during fabrication[24,25]. Later, Wang *et al.*[26] experimentally demonstrated the observation of out-of-plane spin-induced anomalous spin-orbit torques by using the polar magneto-optic Kerr effect. Fu *et al.* recently reported the out-of-plane spin-orbit torques in different single-layer FM through spin-torque ferromagnetic resonance (ST-FMR).[27] However, these observations were indirect measurements of out-of-plane spins and relied on unconventional spin-orbit torques. This has led to some controversy as it is generally believed that spin-orbit torques cannot be generated in a uniform single-layer FM by a uniformly distributed current. Therefore, the direct observation of out-of-plane spins is necessary.

In this work, we directly visualize out-of-plane spin accumulation on the top surface of the single layer Co by helicity-dependent scanning photovoltage microscopy[28-30]. We investigate the Co magnetization direction, current density, and thickness-dependent photovoltages. Then the out-of-plane spin generation efficiency is estimated. Our work broadens the possibility of sputtered and easily fabricated out-of-plane spin sources and the understanding of the out-of-plane spin accumulation.

The Co ($x$ nm)/MgO (1 nm)/SiO$_2$ (2 nm) structures, where $x$ represents the thickness (5, 7.5, 10, 12.5, 15, 17.5, 20 and 22.5 nm) of the Co layer, are deposited onto thermally oxidized Si substrates using magnetron sputtering with a base pressure of < 3 × 10$^{-9}$ Torr. The Co layer is deposited using a DC power of 60 W and a pressure of 3 mTorr. The capping layers, MgO (1 nm)/SiO$_2$ (2 nm), are deposited using an RF power of 120 W and a pressure of 2 mTorr without reactive gas. To ensure accurate thicknesses, the deposition rate for each target is calibrated before deposition. The films (X-ray diffraction results can be found in Fig. S1, Supplementary Information) are patterned into standard Hall bars with a typical lithography method followed by Ar etching. The dimensions of the Hall bar devices are 90 μm in the longitudinal direction and 40 μm in the transverse direction with a Hall bar width of 10 μm. After removing the oxidization capping layer by ion beam etching, the electrode contacts of Ta (5 nm)/Cu (100 nm)/Ta (5 nm) are fabricated by sputtering and lift-off processes. The fabricated devices exhibit similar reversal curves, saturation magnetization ($M_s$) and remanence ratio ($M_r/M_s$) characteristics across different thicknesses (see Fig. S2-S4, Supplementary Information for details).

The experimental setup is depicted in Fig. 1. A laser with a central wavelength of 660 nm serves as the probe for helicity-dependent scanning photovoltage measurements. An attenuator is employed to regulate the laser power to 10 mW. The laser polarization is set to 45° to the axis of the photoelastic modulator (PEM) using a polarizer and half-wave plate. Consequently, the right circular polarized (RCP) light or left circularly polarized (LCP) light can be modulated at a frequency of 50 kHz. The RCP/LCP laser is then focused onto the device using an 80× long working distance microscope objective lens at normal incidence. The area where the laser shines on the device can be adjusted with a motorized *xyz*-piezo stage, aided by a beam splitter and optical imaging module.

When the circularly polarized light is normally shining on the device, RCP and LCP generate electrons with opposite out-of-plane spins. When there is no direct bias current applied along the device, the RCP or LCP light is equally absorbed. The helicity-dependent voltage ($V_\pm = V_{RCP} - V_{LCP}$) is therefore equal to zero, as the RCP and LCP generated photovoltages, $V_{RCP}$ and $V_{LCP}$, are equal. On the other hand, the helicity-dependent voltage is non-zero when the bias current is applied. In this case, the current-induced out-of-plane spins accumulate at the device surface. Due to the magnetic circular dichroism effects, there is variation in RCP and LCP absorption at the surface when the current-induced local spin accumulation exists. The variation in RCP and LCP absorption leads to a difference between $V_{RCP}$ and $V_{LCP}$. As a result, a non-zero $V_\pm$ is produced, which reflects the distribution of local out-of-plane spins. To acquire the $V_{RCP}$ and $V_{LCP}$ with a high sensitivity, a lock-in amplifier is involved while the trigger signal is synchronized with the frequency signal of PEM as shown in Fig. 1.

Figure 2(a-d) illustrates the photovoltage mapping results obtained from a substrate/Co (15 nm)/MgO (1 nm)/SiO$_2$ (2 nm) device with different magnetization initialization directions. The scanning area covers 16 μm × 16 μm with a spatial resolution of 30 pixels × 30 pixels. The black dashed lines in Fig. 2(b,d and f) indicate the device boundaries while the black arrows indicate the bias current (*I*) direction and initial magnetization (**M**) directions. The device in Fig. 2(a) is first saturated along the *x*-axis under a magnetic field of ~500 Oe. During the measurement, the magnetic field is removed. A uniform magnetization distribution within our device in the remanence state can be observed (Fig. S5, Supplementary

Information). When the magnetization direction is parallel to the current direction (**M** // *I* // *x*), a clear surface $V_\pm$ can be observed when 8 mA current is applied in Fig. 2(b). It indicates that out-of-plane spin is generated with a current density ($J_c$) of $5.33 \times 10^6$ A/cm$^2$. However, when **M** is initialized along the −*y*-axis (**M** // −*y*) and the current flows along the *x*-axis (*I* // *x*) as shown in Fig. 2(c), perpendicular to the magnetization, no clear $V_\pm$ signal is observed as depicted in Fig. 2(d). This indicates the absence of out-of-plane spin in this configuration. This result is consistent with previously reported findings for a 32 nm thick Py film using the polar magneto-optic Kerr effect[26]. To further validate the measurements and exclude potential artefacts from the system itself[31], photovoltage $V_\pm$ measurements are also performed on Cu (15 nm)/MgO (1 nm)/SiO$_2$ (2 nm) devices. The device and measurement geometries for Cu devices are the same as for Co devices. Cu is a typical normal metal with negligible spin-orbit coupling, and thus, out-of-plane spins cannot be generated when the current flows through Cu. As expected, no significant $V_\pm$ signal is observed on the surface or edge of the Cu device, as shown in Fig. 2(f).

To investigate whether the generation of out-of-plane spin is induced by currents, we conduct $V_\pm$ measurements with different biases. The scanning area covers 16 μm × 20 μm with a spatial resolution of 32 pixels × 40 pixels. In Fig. 3(c), it can be observed that the helicity-dependent photovoltage $V_\pm$ signal is negligible without any bias applied. The sign of $V_\pm$ reverses when the current direction is reversed. Furthermore, the magnitude of surface $V_\pm$ increases as the current magnitude is increased. These observations provide strong evidence of current-induced out-of-plane spin accumulation at the surface of the 15 nm thick Co device. In comparison, a heavy metal Pt only generates a non-zero $V_\pm$ signal at the edges of the device, as shown in previous studies[29]. This is because the spin Hall effect in Pt predominantly contributes to in-plane spin accumulation at the device surfaces and out-of-plane spin accumulation at the device boundaries. Since our measurement technique is specifically sensitive to out-of-plane spins, a non-zero $V_\pm$ signal appears only at the edges of the Pt device.

Strong photovoltage mapping results shown in Figs. 2 and 3 demonstrate effective out-of-plane spin generation in the 15 nm thick single-layer Co film. To verify the origin of this spin generation, photovoltage measurements were conducted on devices with varying thicknesses of Co. Figure 4(a) presents the current-dependent average surface photovoltage $V_\pm$ with the current normalized to the current density ($J_c$) with representative device thicknesses. The $V_\pm$ displays a linear correlation with the $J_c$ when the thickness of Co is varied between 5 to 20 nm. The efficiency of out-of-plane spin generation can be quantified by the slope of the $V_\pm$ versus $J_c$ relationship, given by the formula $\zeta = \Delta V_\pm/J_c$. Figure 4(b) presents that the $\zeta$ increases from nearly 0.1 to 0.4 µV/($10^6$ A/cm$^2$) with increasing the thickness from 5 to 22.5 nm. The saturation of $\zeta$ beyond Co thickness of 15 nm can be attributed to the laser penetration depth of around 12 nm into the material (details can be found in Supplementary Note 1). This indicates that bulk out-of-plane spins dominate the photovoltage response in the Co films studied. Based on these observations, it can be concluded that in our study, the photovoltage in Co is primarily generated by bulk out-of-plane spins rather than surface effects or interfacial phenomena.

The mechanism behind the generation of out-of-plane spins can be explained through symmetry discussion. In devices where no current is flowing, mirror symmetries along $xy$-, $xz$-, and $yz$-planes ($\sigma_{xy}$, $\sigma_{xz}$, and $\sigma_{yz}$), two-fold rotational symmetries along $x$-, $y$- and $z$-axis ($C_2^x$, $C_2^y$ and $C_2^z$), and center inversion symmetry ($i$) are preserved[26]. However, when the current flows along the $x$-axis of nonmagnetic materials such as Cu, mirror symmetry along the $yz$-plane, and rotation symmetries along the $y$- and $z$-axis are broken. If we assume that the out-of-plane spins can be generated, when the current flows along the $x$-axis as shown in Fig. 5(a), the polar vector current direction remains the same while the axis vector spin direction is reversed after performing the symmetry operation $\sigma_{xz}$. Importantly, it should be noted that under the same conditions, the generation of out-of-plane spins with reversed moments (i.e., up-spins and down-spins) is not permitted. Consequently, the current-induced out-of-plane spin is forbidden in a nonmagnetic metal based on symmetry considerations. This finding is in accordance with the experimental results presented in Fig. 2(f), where negligible out-of-plane spin generation is observed in the Cu device.

In magnetic devices, the symmetries are further diminished by the presence of magnetization and its direction. When the device is magnetized along the *x*-axis, which is parallel to the current direction, only the rotational symmetry along the *x*-axis $C_2^x$ remains intact, while all mirror symmetries are broken. By applying the symmetry operation $C_2^x$, both the polar and axis vectors maintain their original directions, as depicted in Fig. 5(b). This symmetry allows for the generation of out-of-plane spin in response to the current, consistent with the experimental results shown in Fig. 2(b). However, when the device is magnetized along the *y*-axis, which is perpendicular to the current direction, only the mirror symmetry along the *xz*-plane ($\sigma_{xz}$) is preserved. Figure 5(c) illustrates that the generation of out-of-plane spin is not possible, as the spin distributions conflict with each other before and after applying $\sigma_{xz}$. Consequently, the helicity-dependent $V_{\pm}$ signal cannot be observed, as depicted in Fig. 2(d).

In conclusion, our study focused on investigating the current-induced accumulation of out-of-plane spins in a single-layer ferromagnetic Co. We employed the helicity-dependent photovoltage mapping technique to directly visualize the distribution of these out-of-plane spins. Our results demonstrate that the generation of out-of-plane spins occurs when the magnetization is aligned parallel to the direction of the current flow. The magnitude of these spins shows a linear correlation with the current density and exhibits a uniform distribution across the surface of the device. Conversely, when the magnetization is initially oriented perpendicular to the current direction, the generation of out-of-plane spins is suppressed. Furthermore, we observed that the efficiency of the photovoltage increases with increasing thickness of the Co layer. In the case of a 22.5 nm thick Co sample, the estimated photovoltage efficiency is approximately 0.4 µV/(10$^6$ A/cm$^2$). The interlayer exchange coupling between in-plane magnetized Co and PMA layer can be utilized with a spacer layer Ru[32] in order to integrate with a magnetic tunnel junction. These findings highlight the potential of using sputtered spin source materials as a practical and efficient approach for achieving field-free switching in spintronic applications.

## SUPPLEMENTARY MATERIAL

See the supplementary material for the laser penetration depth (Note 1), film characterization results (Fig. S1), and magnetization characterization (Figs. S2 – S5).

## ACKNOWLEDGMENTS

This work was supported by SpOT-LITE program (A*STAR grant, A18A6b0057) through RIE2020 funds, Samsung Electronics Co., Ltd (IO221024-03172-01), and National Research Foundation (NRF) Singapore Investigatorship (NRFI06-2020-0015).

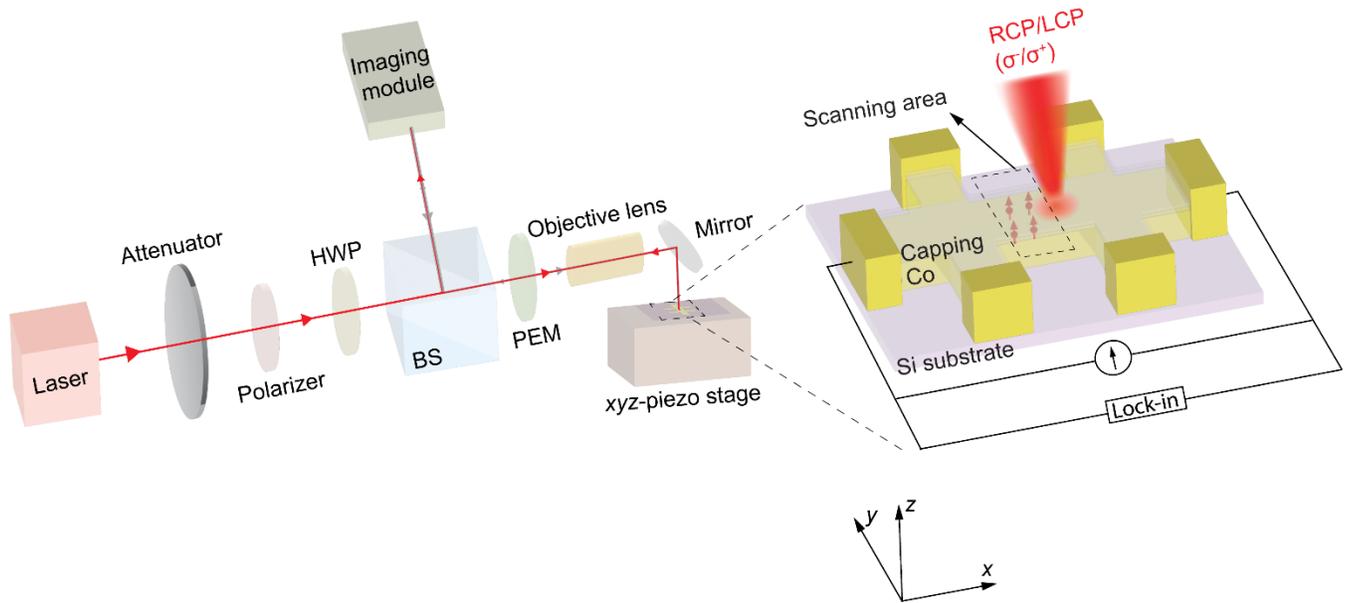

FIG. 1. Schematics of the experiment setup. The focused 660 nm laser is normally incident to the device. An imaging module is utilized to identify the location of the device and laser spot. Two-dimensional scanning is achieved through the movement of the *xyz*-piezo stage while the laser spot remains at rest. HWP, BS, and PEM represent a half wave plate, beam splitter, and photoelastic modulator, respectively. Schematics of the measurement when the current flows along the *x*-axis.

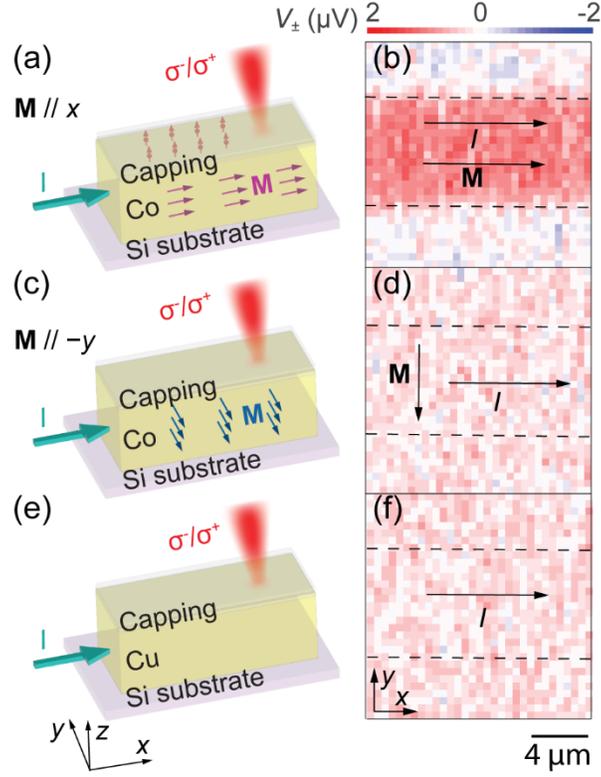

FIG. 2. Illustrations of the $V_{\pm}$ mappings in substrate/Co (15 nm)/MgO (1 nm)/SiO$_2$ (2 nm) device with (a) **M** // *I* // *x*, (c) **M** // −*y* and *I* // *x* and (e) substrate/Cu (15 nm)/MgO (1 nm)/SiO$_2$ (2 nm) device with *I* // *x*. The green, red, purple, and blue arrows represent the current flow, out-of-plane spins, **M** // *x* and **M** // −*y*, respectively. Spatial two-dimensional $V_{\pm}$ mappings when the magnetization is along the (b) *x*-axis (**M** // *x*) and (d) −*y*-axis (**M** // −*y*) with the current of 8 mA along the *x*-axis in Co. (f) Spatial two-dimensional $V_{\pm}$ mappings when a current of 8 mA flows in substrate/Cu (15 nm)/MgO (1 nm)/SiO$_2$ (2 nm) device. Black dashed lines indicate the device boundaries and the black arrows correspond to the current (*I*) and magnetization (**M**) direction.

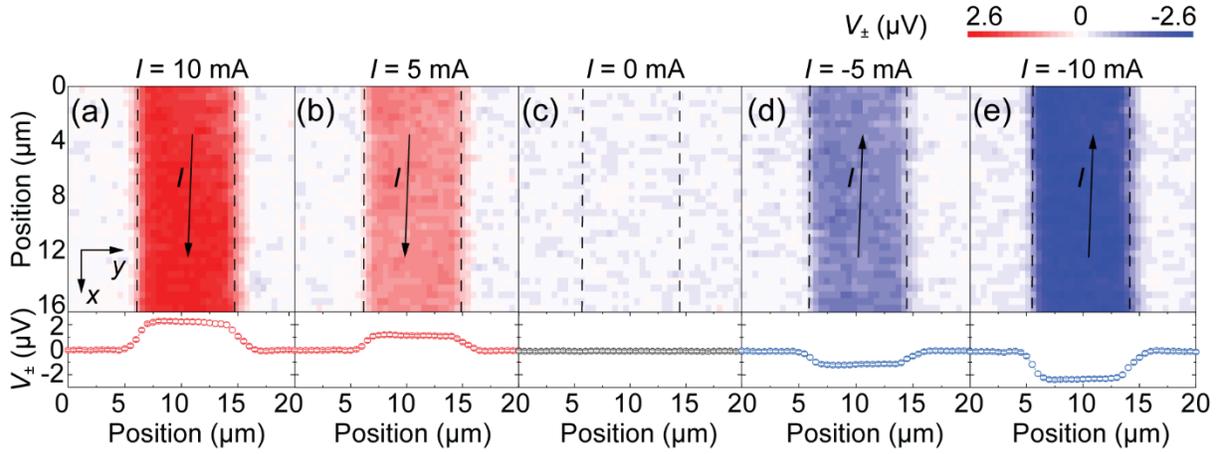

FIG. 3. Spatial two-dimensional $V_\pm$ mappings with the bias current of (a) 10, (b) 5, (c) 0, (d) −5, and (e) −10 mA applied along the *x*-axis in 15 nm thick Co. Black dashed lines indicate the device boundaries and the black solid arrows correspond to the current direction.

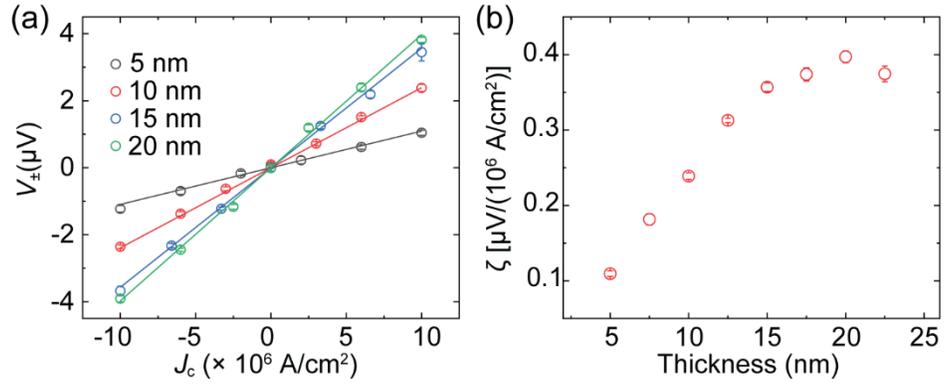

FIG. 4. $V_\pm$ in different Co thicknesses. (a) The current density dependent $V_\pm$ with representative Co thicknesses. (b) The out-of-plane spin generation efficiency $\zeta$.

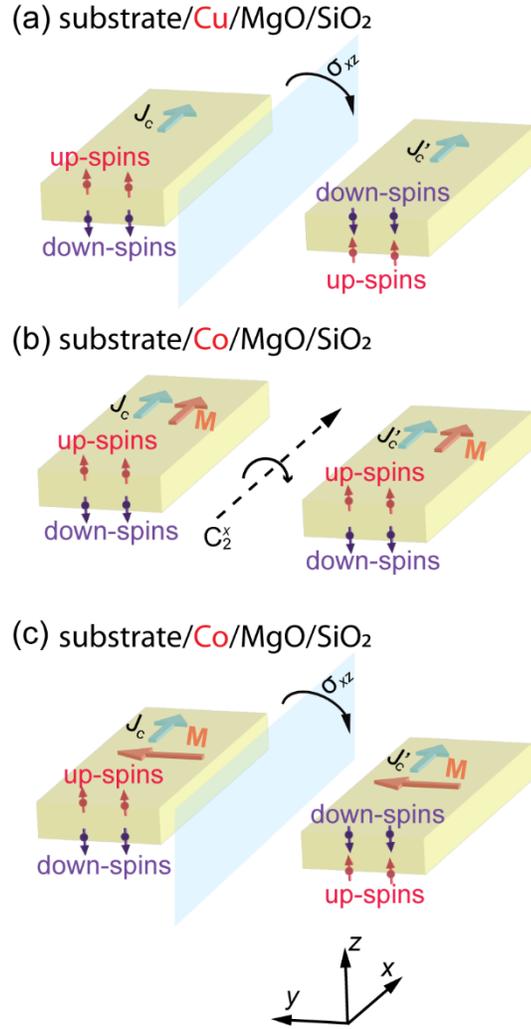

FIG. 5 Symmetry discussion for out-of-plane spin generation. The red and purple arrows with balls correspond to the up- and down-spins, respectively. The current and magnetization direction are indicated with green and red arrows, respectively. (a) In a non-magnetic device, e.g. substrate/Cu/MgO/SiO$_2$, the out-of-plane spin is forbidden by symmetry. (b) In a magnetic device, e.g. substrate/Co/MgO/SiO$_2$, when the magnetization is parallel to the current direction, symmetry allows the out-of-plane spin. (c) In a magnetic device, e.g. substrate/Co/MgO/SiO$_2$, the out-of-plane spin is not allowed when the magnetization is orthogonal to the current direction.